\documentclass[aps,pra,preprint,showpacs]{revtex4}
\usepackage{graphicx}
\usepackage{amsmath}
\usepackage{amsfonts}
\usepackage{amssymb}
\usepackage{dcolumn}  

\newcommand*{\cent}[1]{\multicolumn{1}{c}{$#1$}}
\newcolumntype{w}[1]{D{.}{.}{#1}}
\begin{document}
\preprint{Version 3.1}

\title{Correlated exponential functions in high precision calculations for diatomic molecules}

\author{Krzysztof Pachucki}
\email[]{krp@fuw.edu.pl}

\affiliation{Faculty of Physics, University of Warsaw,
             Ho\.{z}a 69, 00-681 Warsaw, Poland}

\begin{abstract}
Various properties of the general two-center two-electron integral over the 
explicitly correlated exponential function
are analyzed for the potential use in high precision calculations for diatomic molecules.
A compact one dimensional integral representation is found,
which is suited for the numerical evaluation. Together with recurrence relations, 
it makes possible the calculation of the two-center two-electron integral with arbitrary powers
of electron distances. Alternative approach via  
the Taylor series in the internuclear distance is also investigated.
Although numerically slower, it can be used in cases  when recurrences lose stability.
Separate analysis is devoted to molecular integrals with integer powers of interelectronic distances $r_{12}$ 
and the vanishing corresponding nonlinear parameter. Several methods of their evaluation are proposed.
\end{abstract}

\maketitle
\section{Introduction}
In order to achieve high accuracy for the nonrelativistic energy as well as for relativistic 
and quantum electrodynamics (QED) corrections in molecular systems, well optimized
basis functions have to be applied. The explicitly correlated Gaussian (ECG) functions
do not satisfy the cusp condition, therefore their use in the evaluation of higher order 
QED corrections \cite{singlet} is problematic. Nevertheless, the recent calculations
of the leading $O(\alpha^3)$ QED effects performed for H$_2$ \cite{komasa} using ECG functions,
lead to the most accurate to date theoretical predictions of about $10^{-3}$ cm$^{-1}$ 
uncertainty for dissociation energies.
To obtain even more accurate results and to include higher order QED effects
we aim to use the basis of explicitly correlated exponential functions of the form
\begin{equation}
\phi =  e^{-w_1\,r_{12}-u_3\,r_{1A}-u_2\,r_{1B}-w_2\,r_{2A}-w_3\,r_{2B}}
        \,r_{12}^{n_1}\,r_{1A}^{n_2}\,\,r_{1B}^{n_3}\,r_{2A}^{n_4}\,r_{2B}^{n_5}. \label{00}
\end{equation}
The notation for interparticle distances is explained after Eq. (\ref{03}).
The use of such exponential functions in molecular applications is quite limited.
Starting from the pioneering work of Kolos and Roothan \cite{rmp}, the Neumann expansion
of $r_{12}^{-1}$ (at $\omega = 0$) in spherical oblate coordinates
\cite{flammer} has been applied most often,
see the most recent review by Harris in Ref. \cite{harris2} and the collection
of works in Ref. \cite{ozdogan}.
This expansion and the more general one of $e^{-\omega\,r_{12}}\,r_{12}^{-1}$ 
in terms of spheroidal functions \cite{flammer},
has been applied by several authors \cite{kolos, wol, sims, nakatsuji, serra} in their 
accurate calculations for the hydrogen molecule.

An alternative approach to perform integrals with exponential function in Eq. (\ref{00})
has originally been proposed in Ref. \cite{fh}.
Authors have obtained an analytic, although quite complex formula for 
the general four-body integral with exponential functions. By taking the inverse
Laplace transform [see Eq. (\ref{05})] in one of the nonlinear parameters,
one can in principle obtain the general two-center two-electron integral in Eq.~(\ref{01}).
Due to the very complicated analytic structure of the four-body formula, 
this inverse Laplace transform has not been applied so far. In our recent work \cite{rec_h2},
we have reformulated the problem of the calculation of the inverse Laplace transform,
into the solution of some differential equations. 
From these differential equations satisfied by the master integral $f(r)$ (see Eq. (\ref{01}) below),
one derived analytic recursion relations for integrals with positive integer powers of 
interparticle distances, assuming that the nonlinear parameter corresponding to $r_{12}$ vanishes,
i.e. for standard molecular integrals. Using these recursions one obtained
analytic results for integrals with James-Coolidge  and extended Heitler-London basis functions. 
The application of analytic formulas has been
demonstrated by the calculation of Born-Oppenheimer energies for H$_2$ in Ref. \cite{bo_h2}
with accuracy of $10^{-15}$ au, and for HeH$^+$ in Ref. \cite{bo_hehp} with accuracy of about $10^{-12}$ au.
While James-Coolidge and Heitler-London basis functions work very well for the ground states,
they are not equally good for excited states, where arbitrary values of nonlinear
parameters are needed. This is the main subject of this work, to develop a computational technique
for general two-center two-electron integrals, which can be used for the calculation of
nonrelativistic energies, and also for relativistic and QED effects in two-electron diatomic molecule. 
Finally we think that various properties of the general two-center two-body integral derived here, 
can be applied for the calculation of nonrelativistic energies of an arbitrary diatomic molecule.
This subject has recently been pursued also by Lesiuk and Moszy\'nski in Ref. \cite{lesiuk}.
 
The general structure of this work is the following. In Sec. II we derive the differential equation
for the master integral $f(r)$ in Eq. (\ref{01}). In Sec. III we find a solution in terms of one-dimensional integral
over elementary functions and additionally  consider few special cases. In Sec. IV the Taylor series of $f(r)$ 
in the internuclear distance $r$  is derived, and this shows the general analytic properties of the master integral $f$.
In Sec. V integrals with positive powers of interparticle distances are obtained
by differentiation of the master integral with respect to the corresponding nonlinear parameter.
In Sec. VI we consider standard molecular integrals, i.e., with the vanishing nonlinear parameter
corresponding to $r_{12}$. In Sec. VII we present a summary with possible further applications.

\section{The master integral}
We introduce here notation and definitions following the previous work in Ref. \cite{rec_h2},
and obtain the fourth order differential equation which is satisfied by the master integral $f(r)$.
Matrix elements of the nonrelativistic Hamiltonian between functions of the type (\ref{00})
can be expressed in terms of $f(r)$ and its derivatives with respect to nonlinear parameters.
The master two-electron two-center integral $f(r)$ is defined by
\begin{equation}
f(r) = r\,\int \frac{d^3 r_1}{4\,\pi}\,\int \frac{d^3 r_2}{4\,\pi}\,
\frac{e^{-w_1\,r_{12}}}{r_{12}}\,
\frac{e^{-u_3\,r_{1A}}}{r_{1A}}\,
\frac{e^{-u_2\,r_{1B}}}{r_{1B}}\,
\frac{e^{-w_2\,r_{2A}}}{r_{2A}}\,
\frac{e^{-w_3\,r_{2B}}}{r_{2B}}, \label{01}
\end{equation}
and the related class of integrals with the nonnegative integer $n$ is
\begin{equation}
f(r,n) = r\,\int \frac{d^3 r_1}{4\,\pi}\,\int \frac{d^3 r_2}{4\,\pi}\,
\frac{1}{r_{12}^{1-n}}\,
\frac{e^{-u_3\,r_{1A}}}{r_{1A}}\,
\frac{e^{-u_2\,r_{1B}}}{r_{1B}}\,
\frac{e^{-w_2\,r_{2A}}}{r_{2A}}\,
\frac{e^{-w_3\,r_{2B}}}{r_{2B}}, \label{02}
\end{equation}
where $1,2$ are indices of the electrons, $A,B$ that of the nuclei, and $r=r_{AB}$ is the distance between the nuclei.
The notation for nonlinear parameters comes from the general four-body integral $g$ defined by
\begin{equation}
g(u_1) = \int \frac{d^3 \rho_1}{4\,\pi}\,\int \frac{d^3 \rho_2}{4\,\pi}\,\int
\frac{d^3 \rho_3}{4\,\pi} \,\frac{e^{-w_1\,\rho_1-w_2\,\rho_2-w_3\,\rho_3
 -u_1\,\rho_{23}-u_2\,\rho_{31}-u_3\,\rho_{12}}}{
\rho_{23}\,\rho_{31}\,\rho_{12}\,\rho_{1}\,\rho_{2}\,\rho_{3}},\label{03}
\end{equation}
with $\vec\rho_1=\vec r_{12},\,\vec\rho_2=\vec r_{2A},\,\vec\rho_3=\vec r_{2B}$.
Function $g$  is related to $f$ by a Laplace transform, namely
\begin{equation}
g(t) =
\int\frac{d^3r}{4\,\pi}\,f(r)\,\frac{e^{-t\,r}}{r^2} = \int_0^\infty dr\,f(r)\,e^{-t\,r},\label{04}
\end{equation}
and the opposite relation is the inverse Laplace transform
\begin{equation}
f(r) =
\frac{1}{2\,\pi\,i}\,\int_{-i\,\infty+\epsilon}^{i\,\infty+\epsilon}dt\,e^{t\,r}\,g(t).
\label{05}
\end{equation}
It was shown in Ref. \cite{rec_h2}, that the function $g$ satisfies the first order differential equation
in any of it's parameter $\alpha = u_i,\,w_i$:
\begin{equation}
\sigma\,\frac{\partial g}{\partial \alpha } +
\frac{1}{2}\,\frac{\partial\sigma}{\partial \alpha}\,g + P_\alpha = 0\,,
\label{06}
\end{equation}
or equivalently
\begin{equation}
\sqrt{\sigma}\frac{\partial}{\partial \alpha}(\sqrt{\sigma}\,g) +
P_\alpha = 0\,,\label{07}
\end{equation}
with the inhomogeneous term $P_\alpha$  presented in Appendix A,
and $\sigma$ being the sixth order polynomial in six variables
\begin{eqnarray}
\sigma &=&
u_1^2\,u_2^2\,w_3^2  + u_2^2\,u_3^2\,w_1^2  + u_1^2\,u_3^2\,w_2^2 + w_1^2\,w_2^2\,w_3^2
 + u_1^2\,w_1^2\,(u_1^2+w_1^2-u_2^2-u_3^2-w_2^2-w_3^2)
\nonumber \\ &&
 + u_2^2\,w_2^2\,(u_2^2+w_2^2-u_1^2-u_3^2-w_1^2-w_3^2)
 + u_3^2\,w_3^2\,(u_3^2+w_3^2-u_2^2-u_1^2-w_1^2-w_2^2).\label{08}
\end{eqnarray}
For example, Eq.~(\ref{06}) for $\alpha = w_1$ reads
\begin{equation}
\sigma\,\frac{\partial g}{\partial w_1} +
\frac{1}{2}\,\frac{\partial\sigma}{\partial w_1}\,g +
P(w_1,u_1;w_2,u_2;w_3,u_3) = 0\,,\label{09}
\end{equation}
where the inhomogeneous term $P(w_1,u_1;w_2,u_2;w_3,u_3)$ is given by Eq. (\ref{A2}).
The solution of this differential equation is presented in
the ingenious work of Fromm and Hill \cite{fh} by direct
integration of Eq. (\ref{01}) in the momentum representation.
A more compact formula was obtained by Harris \cite{harris}.
We do not present here their results, as its explicit form is quite long
and we will not need it.

To obtain the fundamental differential equation for the master integral $f$,
we use the differential equation (\ref{06}) in variable $t=u_1$
\begin{equation}
\sigma\,\frac{\partial g}{\partial t} +
\frac{1}{2}\,\frac{\partial\sigma}{\partial t}\,g + P(t,w_1;u_3,w_3;w_2,u_2) = 0\,,
\label{10}
\end{equation}
and perform the inverse Laplace transform in $t$.
This differential equation, using the following new parameters
which are adapted to the symmetry of the problem
\begin{equation}
w_2 = w+x,\;\;
w_3 = w-x,\;\;
u_2 = u-y,\;\;
u_3 = u+y,\label{11}
\end{equation}
takes the form
\begin{equation}
\biggl[\sigma_4\,\frac{d^2}{d\,r^2}\,r\,\frac{d^2}{d\,r^2} + \sigma_2\,\frac{d}{d\,r}\,r\,\frac{d}{d\,r} 
+ \sigma_0\,r\biggr]\,f(r)  = F(r), \label{12}
\end{equation}
where
\begin{eqnarray}
\sigma &=&\sigma_0 +t^2\,\sigma_2 + t^4\,\sigma_4, \label{13}\\
\sigma_4 &=& w_1^2, \nonumber \\
\sigma_2 &=& w_1^4\,-2\,w_1^2\,(u^2+w^2+x^2+y^2)+16\,u\,w\,x\,y \nonumber \\
         &=& w_1^4 + w_1^2\,\sigma_{22} + \sigma_{20},\nonumber \\
\sigma_0 &=& w_1^2\,(u + w - x - y)\,(u - w + x - y)\,(u - w - x + y)\,(u + w + x + y) \nonumber \\&&
        + 16\,(w\,x - u\,y)\,(u\,x - w\,y)\,(u\,w - x\,y) 
\nonumber \\
&=&  w_1^2\,\sigma_{02} + \sigma_{00}, \nonumber
\end{eqnarray}
and $F(r) = F_{u_1}(r)$ is presented in Appendix A.
The differential Eq. (\ref{12}) is supplemented by the boundary conditions,
namely $f(r)$ vanishes at small and large $r$. 

Taking the occasion,  we present here, for later use,
a differential equation which is obtained from the inverse Laplace transform of Eq. (\ref{09})
\begin{equation}
\Bigl(\frac{1}{2}\,\frac{\partial\sigma_0}{\partial w_1} + \sigma_0\,\frac{\partial }{\partial w_1}\Bigr)\,f(r) 
+\Bigl(\frac{1}{2}\,\frac{\partial\sigma_2}{\partial w_1} + \sigma_2\,\frac{\partial}{\partial w_1}\Bigr)\,f''(r) 
+ \Bigl(w_1 + w_1^2\,\frac{\partial}{\partial w_1} \Bigr)\,f^{(4)}(r)
= -F_{w_1}(r), \label{14}
\end{equation}
where $F_{w_1}(r)$ is given in Appendix A, and that for an arbitrary parameter $\alpha = u,w,x,y$
\begin{equation}
\Bigl(\frac{1}{2}\,\frac{\partial\sigma_0}{\partial \alpha} + \sigma_0\,\frac{\partial }{\partial \alpha}\Bigr)\,f(r) 
+\Bigl(\frac{1}{2}\,\frac{\partial\sigma_2}{\partial \alpha} + \sigma_2\,\frac{\partial}{\partial \alpha}\Bigr)\,f''(r) 
+ w_1^2\,\frac{\partial}{\partial \alpha}\,f^{(4)}(r)
= -F_\alpha(r). \label{15}
\end{equation}

\section{Poisson representation} 
We derive here the integral representation for solutions of the homogeneous and the inhomogeneous
differential equation (\ref{12}), which is the analog of the Poisson representation for Bessel functions \cite{nist}.
$\sigma$ is a quadratic polynomial in $t^2$, so it has four zeros $\pm t_a$, $\pm t_b$.
Assuming $t_b > t_a > 0$, the four solutions of the homogeneous equation
\begin{equation}
\biggl[\sigma_4\,\frac{d^2}{d\,r^2}\,r\,\frac{d^2}{d\,r^2} 
+ \sigma_2\,\frac{d}{d\,r}\,r\,\frac{d}{d\,r} + \sigma_0\,r\biggr]\,f(r)  = 0,
\label{16}
\end{equation}
are
\begin{equation}
\int_{-\infty}^{-t_b}\,\frac{e^{t\,r}}{\sqrt{\sigma(t)}}\,dt,\;\;
\int_{-t_b}^{-t_a}\,\frac{e^{t\,r}}{\sqrt{\sigma(t)}}\,dt,\;\;
\int_{-t_a}^{t_a}\,\frac{e^{t\,r}}{\sqrt{\sigma(t)}}\,dt,\;\;
\int_{t_a}^{t_b}\,\frac{e^{t\,r}}{\sqrt{\sigma(t)}}\,dt\,.
\label{17}
\end{equation}
Let us prove it, as an example, for the last term
\begin{align}
&\biggl[\sigma_4\,\frac{d^2}{d\,r^2}\,r\,\frac{d^2}{d\,r^2} + \sigma_2\,\frac{d}{d\,r}\,r\,\frac{d}{d\,r} + \sigma_0\,r\biggr]
\int_{t_a}^{t_b} dt\,\frac{e^{t\,r}}{\sqrt{\sigma(t)}} \nonumber \\
=&\int_{t_a}^{t_b} \frac{dt}{\sqrt{\sigma(t)}}\,
\biggl[\sigma_4\,\frac{d^2}{d\,r^2}\,r\,\frac{d^2}{d\,r^2} + \sigma_2\,\frac{d}{d\,r}\,r\,\frac{d}{d\,r} + \sigma_0\,r\biggr]\,
e^{t\,r}\nonumber \\
=&\int_{t_a}^{t_b} \frac{dt}{\sqrt{\sigma(t)}}\,
\biggl[\sigma_4\,t^2\,\frac{d}{dt}\,t^2 + \sigma_2\,t\,\frac{d}{dt}\,t + \sigma_0\,\frac{d}{dt}\biggr]\,e^{t\,r}\nonumber \\
=&\,\frac{1}{2}\,\int_{t_a}^{t_b} \frac{dt}{\sqrt{\sigma(t)}}\,
\biggl[\sigma(t)\,\frac{d}{dt} + \frac{d}{dt}\,\sigma(t)\biggr]\,e^{t\,r}, \label{18}
\end{align}
where the last equation holds because $\sigma(t) = \sigma_0 + \sigma_2\,t^2+\sigma_4\,t^4$.
One integrates by parts, boundary terms vanish because $\sigma(t_a) =
\sigma(t_b) = 0$, and one obtains
\begin{align}
=&\,\frac{1}{2}\,\int_{t_a}^{t_b} dt\,e^{t\,r}\,
\biggl[\frac{d}{dt}\,\sqrt{\sigma(t)} + \sigma(t)\,\frac{d}{dt}\,\frac{1}{\sqrt{\sigma(t)}}\biggr]
\nonumber \\
=& \,\frac{1}{2}\,\int_{t_a}^{t_b} dt\,e^{t\,r}\,\frac{1}{2\,\sqrt{\sigma(t)}}\,
(1-1) = 0. \label{19}
\end{align}

The solution of the inhomogeneous fundamental differential equation (\ref{12})
is obtained by noting that the integration contour
in the inverse Laplace transform can be deformed to encircle all branch cuts on 
the left side of the complex plain
\begin{equation}
f(r) = \frac{1}{2\,\pi\,i}\,\int_{0}^{-\infty}dt\,e^{t\,r}\,\bigl[g(t+i\,\epsilon) - g(t-i\,\epsilon)\bigr].
\label{20}
\end{equation}
The analytic properties of $g$ have been analyzed by Fromm and Hill in \cite{fh}. The function
$g(t)$ has four overlapping branch cuts, similarly to the function $P$, in variable $t$
on the negative real axis starting at
\begin{eqnarray}
-t_1 =& u_3 + w_2       &= u + y + w + x,  \nonumber \\
-t_2 =& u_2 + w_3       &= u - y + w - x,  \nonumber \\
-t_3 =& u_3 + w_1 + w_3 &= u + y + w - x + w_1, \nonumber \\  
-t_4 =& u_2 + w_1 + w_2 &= u - y + w + x + w_1, \label{21}
\end{eqnarray}
correspondingly. The fact that all $t_i$ with $i=1,2,3,4$ are not positive comes from the requirement
that the integral in Eq. (\ref{01}) is finite for positive values of $u_i$ and $w_i$. 
Fromm and Hill found the imaginary part
of $g$ for the particular ordering of $t_i$. Their result is generalized here
to an arbitrary ordering of $t_i$ and takes the form
\begin{eqnarray}
\frac{1}{2\,\pi\,i}\,\bigl[g(t+i\,\epsilon) - g(t-i\,\epsilon)\bigr] &=& 
\frac{1}{2\,\sqrt{\sigma}}\bigl[
\theta(t_1-t)\,\ln|\beta_{0,0}| + 
\theta(t_2-t)\,\ln|\beta_{3,3}| 
\nonumber \\ &&
-\theta(t_3-t)\,\ln|\beta_{3,1}\,\beta_{3,3}|
-\theta(t_4-t)\,\ln|\beta_{0,1}\,\beta_{0,0}|\bigr], \label{22}
\end{eqnarray}
where
\begin{equation}
\beta_{i,j} = \frac{\sqrt{\sigma}-\gamma_{i,j}}{\sqrt{\sigma}+\gamma_{i,j}}, \label{23}
\end{equation}
and (cf. Ref.\cite{fh} with $a_1 = u_3, a_2 = w_2,  a_3 = t, a_{23}= w_3, a_{31}= u_2, a_{12}= w_1$)
\begin{eqnarray}
\gamma_{0,0} &=& 2\,u_2\,w_1\,w_3 + (u_2^2 - u_3^2 + w_1^2)\,w_3 + w_1\,(-t^2 + u_2^2 + w_3^2) + u_2\,(w_1^2 - w_2^2 + w_3^2),
\nonumber\\
\gamma_{3,3} &=& 2\,u_3\,w_1\,w_2 + (-u_2^2 + u_3^2 + w_1^2)\,w_2 + w_1\,(-t^2 + u_3^2 + w_2^2) + u_3\,(w_1^2 + w_2^2 - w_3^2),
\nonumber\\
\gamma_{0,1} &=& -2\,t\,w_2\,w_3 - (t^2 - u_3^2 + w_2^2)\,w_3 + w_2\,(t^2 - u_2^2 + w_3^2) + t\,(-w_1^2 + w_2^2 + w_3^2),
\nonumber\\
\gamma_{3,1} &=& -2\,t\,w_2\,w_3 + (t^2 - u_3^2 + w_2^2)\,w_3 - w_2\,(t^2 - u_2^2 + w_3^2) + t\,(-w_1^2 + w_2^2 + w_3^2).
\label{24}
\end{eqnarray}
There is some arbitrariness in the form of Eq. (\ref{22})
as products of $\beta$ can be expressed in several ways \cite{fh,harris}
\begin{eqnarray}
\beta_{\mu,\mu}\, \beta_{\mu,\nu} &=& \beta_{\nu,\nu}\,\beta_{\nu,\mu} = \beta_{\rho,\rho}\,\beta_{\rho,\sigma}, 
\nonumber \\
\beta_{\mu,\sigma}\,\beta_{\nu,\sigma}\,\beta_{\rho,\sigma} &=& \beta_{\sigma,\sigma},
\label{25}
\end{eqnarray}
for $\{\mu,\nu,\rho,\sigma\}$ being an arbitrary permutation of $\{0,1,2,3\}$.
One can verify that ${\rm Im}(g)$ in Eq. (\ref{22}), satisfies the differential equation (\ref{10}) with ${\rm Im}(P)$ 
as an inhomogeneous term.
As a result, $f(r)$ can be expressed in terms of the one-dimensional integral
\begin{eqnarray}
f(r) &=& 
\int_{0}^{-\infty}dt\,e^{t\,r}\,
\frac{1}{2\,\sqrt{\sigma}}\bigl[
\theta(t_1-t)\,\ln|\beta_{0,0}| + 
\theta(t_2-t)\,\ln|\beta_{3,3}| 
\nonumber \\ &&\hspace*{12ex}
-\theta(t_3-t)\,\ln|\beta_{3,1}\,\beta_{3,3}|
-\theta(t_4-t)\,\ln|\beta_{0,1}\,\beta_{0,0}|\bigr].
\label{26}
\end{eqnarray}
This is a main result obtained in this work. All properties of the master integral $f$
can in principle be obtained from this integral representation. 
Some of them, however, can be more easily obtained from the differential equation (\ref{12}). 
The rest of the paper will be devoted to various properties
and various ways of numerical calculation of $f(r)$ in Eq. (\ref{26})
and its derivatives with respect to nonlinear parameters. 
This form of Eq. (\ref{26}) is suited for the direct numerical quadrature, once we know 
all the singularities on the integration path, and this is presented in Appendix B.

It would be interesting to investigate various expansions of $f(r)$ in Eq. (\ref{26}),
in particular the expansion around $w_1=0$ is studied in Sec. VI.
Here, we obtain the master integral $f$ at $w_1=0$ in a simple form
which is convenient for the numerical evaluation. The $\sigma$ polynomial reads 
\begin{eqnarray}
\sigma|_{w_1=0} &=&
(u_2^2 - u_3^2 + w_2^2 - w_3^2)\,(u_2^2\,w_2^2 - u_3^2\,w_3^2)
- t^2\,(u_2^2 - u_3^2)\,(w_2^2 - w_3^2)\nonumber \\ &=&
16\,(w\,x - u\,y) (u\,x - w\,y) (u\,w - x\,y) + 16\,t^2\,u\,w\,x\,y,
\label{27}
\end{eqnarray}
and the differential equation becomes \cite{rec_h2}
\begin{equation}
 \biggl[16\,u\,w\,x\,y\,\frac{d}{d\,r}\,r\,\frac{d}{d\,r} 
+ 16\,r\,(w\,x - u\,y)\,(u\,x - w\,y)\,(u\,w - x\,y)\biggr]f(r) = F(r)\bigr|_{w_1=0},
\label{28}
\end{equation}
where the inhomogeneous term $F(r)$ using Eq.(\ref{A5}) is 
\begin{eqnarray}
F(r)\bigr|_{w_1=0} &=&
 2\,(u\,w\,x-u\,w\,y-u\,x\,y+w\,x\,y)\,F_{1-}
+2\,(u\,w\,x-u\,w\,y+u\,x\,y-w\,x\,y)\,F_{2-}
\nonumber\\ &&
-2 (u\,w\,x+u\,w\,y+u\,x\,y+w\,x\,y)\,F_{3-}
+2\,(u\,w\,x+u\,w\,y-u\,x\,y-w\,x\,y)\,F_{4-},
\nonumber \\
\label{29}
\end{eqnarray}
and where
\begin{eqnarray}
F_{1\pm} &=& \text{Ei}(-2\,r\,u)\, e^{r\,(u-w-x+y)} \pm \text{Ei}(-2\,r\,w)\, e^{r\,(-u+w+x-y)},\nonumber \\
F_{2\pm} &=& \text{Ei}(-2\,r\,w)\, e^{r\,(-u+w-x+y)} \pm \text{Ei}(-2\,r\,u)\, e^{r\,(u-w+x-y)},\nonumber \\
F_{3\pm} &=& \biggl[\text{Ei}(2\,r\,x)+\text{Ei}(2\,r\,y)-\text{Ei}(2\,r\,(x+y))
+\ln\left(\frac{u\,w\,(x+y)}{x\,y\,(u+w)}\right)\biggr]\,e^{-r\,(u+w+x+y)}
\nonumber \\ &&
   \pm \text{Ei}(-2\,r\,(u+w))\, e^{r\,(u+w+x+y)},\nonumber \\
F_{4\pm} &=& \biggl[\text{Ei}(-2\,r\,x)+\text{Ei}(-2\,r\,y)-\text{Ei}(-2\,r\,(x+y))
+\ln\left(\frac{u\,w\,(x+y)}{x\,y\,(u+w)}\right)\biggr]\,e^{-r\,(u+w-x-y)}
\nonumber \\ &&
   \pm \text{Ei}(-2\,r\,(u+w))\, e^{r\,(u+w-x-y)},
\label{30}
\end{eqnarray}
with Ei being the exponential integral function.
The solution of this differential equation (assuming $p^2>0$)
in terms of solutions of the homogeneous equation was presented in Ref. \cite{rec_h2}.
\begin{equation}
f(r) = -\frac{1}{16\,u\,w\,x\,y}\biggl[
I_0(p\,r)\,\int_r^\infty dr'\,F(r')\,K_0(p\,r') + K_0(p\,r)\,\int_0^r dr'\,F(r')\,I_0(p\,r')\biggr],
\label{31}
\end{equation}
where 
\begin{equation}
p^2 = \frac{(w\,x - u\,y)\,(w\,y - u\,x)\,(u\,w - x\,y)}{u\,w\,x\,y},
\label{32}
\end{equation}
and $I_0$, $K_0$ are modified Bessel functions.
A more convenient, however is the integral form of Eq. (\ref{26})
\begin{eqnarray}
f(r) &=& 
\biggl(\int_{-u-w-x-y}^{-\infty} \ln|\beta_{0,0}| + \int_{-u-w+x+y}^{-\infty} \ln|\beta_{3,3}|
\nonumber \\ &&
-\int_{-u-w+x-y}^{-\infty} \ln|\beta_{3,1}\,\beta_{3,3}| -\int_{-u-w-x+y}^{-\infty} \ln|\beta_{0,1}\,\beta_{0,0}|\biggr)\,
\frac{e^{t\,r}}{2\,\sqrt{\sigma}}\,dt,\nonumber \\
\label{33}
\end{eqnarray}
with the following $\gamma_{i,j}$ coefficients
\begin{eqnarray}
\gamma_{0,0} &=& -4\,(u\,w\,x+u\,w\,y-u\,x\,y-w\,x\,y),
\nonumber\\
\gamma_{3,3} &=& 4\,(u\,w\,x+u\,w\,y+u\,x\,y+w\,x\,y),
\nonumber\\
\Gamma(\beta_{3,1}\,\beta_{3,3}) &=& -16\,u\,w\,x\,y/(t + u + w - x + y) + 4\,(u\,w\,x - u\,w\,y + u\,x\,y + w\,x\,y),
\nonumber\\
\Gamma(\beta_{0,1}\,\beta_{0,0}) &=& -16\,u\,w\,x\,y/(t + u + w + x - y) - 4\,(u\,w\,x - u\,w\,y - u\,x\,y - w\,x\,y),
\label{34}
\end{eqnarray}
%
%
where $\Gamma$ is defined in Eq. (\ref{B4}). 
In the particular case of exchange integral $w=u$, $x=y$, $\sigma_{00}$ vanishes and the master
integral becomes
\begin{eqnarray}
f(r) &=& \frac{1}{8\,w\,x}\,\biggl(
  \int_{-2\,(w+x)}^{-\infty}dt\, \ln\biggl|\frac{t+2\,(w-x)}{t-2\,(w-x)}\biggr|\,\frac{e^{t\,r}}{t}
\nonumber \\ &&
+ \int_{-2\,(w-x)}^{-\infty}dt\, \ln\biggl|\frac{t-2\,(w+x)}{t+2\,(w+x)}\biggr|\,\frac{e^{t\,r}}{t}
\nonumber \\ &&
-2\,\int_{-2\,w}^{-\infty}dt\, \ln\biggl|\frac{t+2\,(w-x)}{t+2\,(w+x)}\biggr|\,\frac{e^{t\,r}}{t}
\biggr).
\label{34p}
\end{eqnarray}
An equivalent integral representation, obtained from the differential equation (\ref{28})
was presented in Ref. \cite{rec_h2}, but this form is more convenient for the numerical evaluation.

\section{Taylor expansion in $r$}
Here, we find the Taylor expansion in the internuclear distance $r$ of the master integral $f$. 
As it was noticed by Fromm and Hill in \cite{fh} this expansion
is absolutely convergent, therefore it is another way to calculate the master integral
as well as its derivatives with respect to $u_i$ and $w_i$.
Following Ref. \cite{fh},  we find the initial terms and the recurrence relation
for subsequent terms of the Taylor series.

The expansion in small $r$ of the master integral has the form
\begin{equation}
f(r) = \sum_{k=1}^\infty r^k\,\bigl[f^{(1)}_k\,\ln(r) + f^{(2)}_k\bigr].
\label{35}
\end{equation}
The leading term, can be obtained by noting that the limit $r\rightarrow 0$
corresponds to the one-center integral \cite{fh}
\begin{align}
&\frac{1}{16\,\pi^2}\int d^3r_1\int d^3r_2\frac{e^{-\alpha\,r_1-\beta\,r_2-\gamma\,r_{12}}}{r_1^2\,r_2^2\,r_{12}}
\nonumber\\=&
\frac{1}{2\,\gamma}\,\biggl[
\frac{\pi^2}{6}+\frac{1}{2}\ln^2\biggl(\frac{\alpha+\gamma}{\beta+\gamma}\biggr)+
{\rm L}_2\biggl(1-\frac{\alpha+\beta}{\alpha+\gamma}\biggr) +
{\rm L}_2\biggl(1-\frac{\beta+\alpha}{\beta+\gamma}\biggr)\biggr],
\label{36}
\end{align}
where L$_2$ is the dilogarthmic function \cite{nist}. Let us introduce 
for the later use the following $X_i$ symbols
\begin{eqnarray}
X_0 &=& \frac{1}{2\,w_1}\,\biggl[
\frac{\pi^2}{6}+\frac{1}{2}\ln^2\biggl(\frac{2\,u+w_1}{2\,w+w_1}\biggr)+
{\rm L}_2\biggl(1-\frac{2\,(u+w)}{2\,u+w_1}\biggr) +
{\rm L}_2\biggl(1-\frac{2\,(u+w)}{2\,w+w_1}\biggr)\biggr],\nonumber \\
X_1 &=& \ln\biggl(\frac{2\,u+w_1}{2\,(u+w)}\biggr),\nonumber \\
X_2 &=& \ln\biggl(\frac{2\,w+w_1}{2\,(u+w)}\biggr),\nonumber \\
X_r &=& \frac{1}{2}\,\ln[r^2\,(2\,u+w_1)(2\,w+w_1)]+\gamma_{\rm E}.
\label{37}
\end{eqnarray}
where $\gamma_{\rm E}$ is the Euler constant.
The leading terms in small $r$ expansion using Eq. (\ref{36}) are
\begin{eqnarray}
f^{(1)}_1 &=& 0,\nonumber \\
f^{(2)}_1 &=& X_0.
\label{38}
\end{eqnarray}
and $f^{(i)}_0 = f^{(i)}_{-1}=0$.
The next terms of this expansion $f^{(1)}_2, f^{(2)}_2$ are obtained from 
the large $t$ asymptotics of the function $g$
\begin{equation}
g = \frac{X_0}{t^2} + \frac{g^{(1)}\,\ln t + g^{(2)}}{t^3} + o\bigl(t^{-4}\bigr).
\label{39}
\end{equation}
Since $g(t)$ satisfies the differential equation (\ref{10}) with the inhomogenous
term $P(t) = P(t,w_1,u_3,w_3,w_2,u_2)$
\begin{equation}
P(t) = w_1^2\,\bigl[2  + \ln(u_2 + u_3 + w_1) + \ln(w_1 + w_2 + w_3)- 2\,\ln(t)\bigr] + o\bigl(t^{-1}\bigr),
\label{40}
\end{equation}
the $1/t^3$ coefficients are
\begin{eqnarray}
g^{(1)} &=& -2, \nonumber \\
g^{(2)} &=& \ln(u_2 + u_3 + w_1) + \ln(w_1 + w_2 + w_3).
\label{41}
\end{eqnarray}
As noticed by Fromm and Hill \cite{fh}, they are related to $f^{(i)}_2$ coefficients by
\begin{eqnarray}
f^{(1)}_2 &=& -\frac{g^{(1)}}{2},\nonumber \\
f^{(2)}_2 &=& \frac{\psi(3)\,g^{(1)}+g^{(2)}}{2},
\label{42}
\end{eqnarray}
where $\psi$ is the Euler $\psi$-function \cite{nist} and $\psi(3) = -\gamma_{\rm E} + 1 + 1/2$.
The next terms in $r$ expansion can be obtained from the fundamental differential equation (\ref{12}).
As the inhomogeneous term $F(r)$ has a similar expansion
\begin{eqnarray}
F(r) = \sum_{k=-1}^\infty r^k\,\bigl(F^{(1)}_k\,\ln(r) + F^{(2)}_k\bigr),
\label{43}
\end{eqnarray}
the recurrence relations are the following
\begin{eqnarray}
f^{(1)}_{k+1} &=& \frac{F^{(1)}_{k-2}-\sigma_0\,f^{(1)}_{k-3}-\sigma_2\,(k-1)^2\,f^{(1)}_{k-1}}{\sigma_4\,(k+1)\,k^2\,(k-1)}, 
\label{44}\\
f^{(2)}_{k+1} &=& \frac{F^{(2)}_{k-2}-\sigma_0\,f^{(2)}_{k-3}-\sigma_2\,(k-1)^2\,f^{(2)}_{k-1}
-\sigma_2\,2\,(k-1)\,f^{(1)}_{k-1} - \sigma_4\,2\,k\,[2\,(k+1)\,(k-1)+1]\,f^{(1)}_{k+1}}{\sigma_4\,(k+1)\,k^2\,(k-1)},
\nonumber 
\end{eqnarray}
The resulting first terms in the expansion of the master integral $f$ in powers of $r$ are
\begin{eqnarray}
f(r) &=& r\,X_0 + r^2\,\Bigl(X_r-\frac{3}{2}\Bigr)
\nonumber \\ &&
+ r^3\,\biggl[-\frac{w_1^2\,X_0}{12} - \frac{w_1}{4}+
\frac{(u^2+w^2+x^2+y^2)\,X_0 - w\,X_1 - u\,X_2 - 2\,(u+w)}{6} \nonumber \\ &&\hspace*{4ex}
+ \frac{x\,y}{3\,w_1} - \frac{2\,x\,y\,(2\,u\,w\,X_0 + u\,X_1 + w\,X_2)}{3\,w_1^2}
\biggr]  
\nonumber \\ &&
+ r^4\,\biggl[\frac{7}{54}\,w_1^2 + \frac{w_1\,(u+w)}{12} + \frac{3\,u\,w - 3\,x\,y - 11\,u^2  - 11\,w^2 - 9\,x^2  - 9\,y^2}{36} 
\nonumber \\&&\hspace*{4ex}
+ X_r\,\biggl(- \frac{w_1^2}{18} + \frac{u^2 + w^2  + x^2 + y^2}{6} \biggr)\biggr] + O(r^5).
\label{45}
\end{eqnarray}

Let's consider now the special case of $w_1 = 0$. The recurrence relation takes the form:
\begin{eqnarray}
f^{(1)}_{k+1} &=& \frac{F^{(1)}_{k}-\sigma_{00}\,f^{(1)}_{k-1}}{\sigma_{20}\,(k+1)^2}, \nonumber \\
f^{(2)}_{k+1} &=& \frac{F^{(2)}_{k}-\sigma_{00}\,f^{(2)}_{k-1}}{\sigma_{20}\,(k+1)^2} - \frac{2}{k+1}\,f^{(1)}_{k+1},
\label{46}
\end{eqnarray}
and the first terms in $r$-expansion are 
\begin{eqnarray}
f(r)|_{w_1=0} &=& 
r\,\biggl[\frac{1}{2\,u}\,\ln\Bigl(1 + \frac{u}{w}\Bigr) + \frac{1}{2\,w}\,\ln\Bigl(1 + \frac{w}{u}\Bigr)\biggr]
+ r^2\,\biggl[-\frac{3}{2} + \gamma + \frac{1}{2}\,\ln(4\,r^2\,u\,w)\biggr]
\nonumber \\ &&
+ r^3\,\biggl[-\frac{u}{3} - \frac{w}{3} + \frac{x\,y}{18\,u} + \frac{x\,y}{18\,w} 
        + \biggl(\frac{u}{4} + \frac{w^2+x^2+y^2}{12\,u} - \frac{w\,x\,y}{18\,u^2}\biggr)
          \,\ln\biggl(1 + \frac{u}{w}\biggr) 
\nonumber \\ && \hspace*{4ex}
        + \biggl(\frac{w}{4} + \frac{u^2+x^2+y^2}{12\,w} - \frac{u\,x\,y}{18\,w^2}\biggr)
           \,\ln\biggl(1 + \frac{w}{u}\biggr)\biggr]
\nonumber \\ &&
+ r^4\,\biggl[\frac{u\,w-x\,y}{12} - \frac{w^2+u^2}{18} + \frac{1}{6}\,(u^2 + w^2 + x^2 + y^2)
              \,\biggl(-\frac{3}{2} + \gamma + \frac{1}{2}\,\ln(4\,r^2\,u\,w)\biggr)\biggr] 
\nonumber \\ &&+ O(r^5).
\label{47}
\end{eqnarray}
In order to obtain the Taylor expansion of $f(r,n)$  
\begin{equation}
f(r,n) = \sum_{k=1}^\infty r^k\,\bigl[f^{(1)}_k(n)\,\ln(r) + f^{(2)}_k(n)\bigr],
\label{48}
\end{equation}
one expands  Eq. (\ref{14}) in power series in $r$ and obtains the following recursions
for coefficients
\begin{eqnarray}
f^{(1)}_{k+2}(n) &=& \frac{-1}{\sigma_{20}\,(k+2)\,(k+1)}\,\biggl[
\sigma_{00}\,f_k^{(1)}(n) + \sigma_{02}\,(n-1)^2\,f_k^{(1)}(n-2) 
\nonumber \\&&
+ \sigma_{22}\,(n-1)^2\,(k+2)\,(k+1)\,f_{k+2}^{(1)}(n-2)
\nonumber \\&&
+(n-1)^2\,(k+4)\,(k+3)\,(k+2)\,(k+1)\,f_{k+4}^{(1)}(n-2) 
\nonumber \\&&
+(n-3)\,(n-2)^2\,(n-1)\,(k+2)\,(k+1)\,f_{k+2}^{(1)}(n-4)
\biggr]\,, \label{49} \\
f^{(2)}_{k+2}(n) &=&  \frac{1}{\sigma_{20}\,(k+2)\,(k+1)}\,\biggl[
F_{w_1,k}(n-1)
-\sigma_{20}\,(2\,k+3)\,f_{k+2}^{(1)}(n)
\nonumber \\&&
-\sigma_{00}\,f_k^{(2)}(n) 
- \sigma_{02}\,(n-1)^2\,f_k^{(2)}(n-2) 
\nonumber \\&&
- \sigma_{22}\,(n-1)^2\,\Bigl[(k+2)\,(k+1)\,f_{k+2}^{(2)}(n-2) + (2\,k+3)\,f_{k+2}^{(1)}(n-2)\Bigr]
\nonumber \\&&
-(n-3)\,(n-2)^2\,(n-1)\,\Bigl[(k+2)\,(k+1)\,f_{k+2}^{(2)}(n-4) + (2\,k+3)\,f_{k+2}^{(1)}(n-4)\Bigr]
\nonumber \\&&
-(n-1)^2\,\Bigl[(k+4)\,(k+3)\,(k+2)\,(k+1)\,f_{k+4}^{(2)}(n-2) 
\nonumber \\&&
\hspace*{10ex} + 2\,(2\,k+5)\,(k^2+5\,k+5)\,f_{k+4}^{(1)}(n-2)\Bigr]\biggr],
\label{50}
\end{eqnarray}
where $f_k^{(i)}(0) = f_k^{(i)}$. The usefulness of the calculation  of $f(r)$ and $f(r,n)$ 
via Taylor series in $r$ and the above recursions needs to be verified numerically.
Nevertheless,
the expansion terms say a lot about analytic properties of $f$, as a function of $w_1, x, y, u$ and $w$.
In particular, the expansion in Eq. (\ref{45}) suggests that $f$ is an entire function 
in $x,y$ at fixed $r,\,u,\,w,\,w_1$.

\section{General recursion relations}
In order to use the exponentially correlated basis in molecular calculations, one has to be able
to calculate derivatives of the master integral $f$ with respect to nonlinear parameters.
Differential equations (\ref{12},\ref{14}) and (\ref{15}) are used below, to express
the derivative of $f(r)$ with respect to an arbitrary parameter $\alpha = w_1, x,y,u,w$
in terms of $f(r)$, $f'(r)$, $f''(r)$, and $f'''(r)$. This is done as follows.
From  Eq. (\ref{12}), the fourth derivative $f^{(4)}(r)$ can be expressed in terms of 
lower derivatives of $f(r)$ and we use this in Eqs. (\ref{14}) and (\ref{15}) to eliminate
fourthth derivative. Next, differentiate these equations with respect to $r$ and again eliminate fourth derivative.
Doing this differentiation and elimination three times, one obtains four equations for four unknowns:
the derivative of $f$ with respect to $w_1$, and its three further derivatives with respect to $r$.
The solution of these linear equations for $\alpha = w_1$ is
\begin{align}
2\,\delta\,\sigma_0\,\frac{\partial f(r)}{\partial w_1}= &
\,r\,\delta\,\frac{\partial\sigma_0}{\partial w_1}\,f'(r) 
+ w_1\,\biggl(-2\,\sigma_0\,\sigma_2 - w_1\,\sigma_2\,\frac{\partial\sigma_0}{\partial w_1}
              + 2\,w_1\,\sigma_0\,\frac{\partial\sigma_2}{\partial w_1}\biggr)\,[f''(r) + r\,f'''(r)]
\nonumber \\ & 
- \sigma_0\,\biggl(4\,w_1\,\sigma_0 + 2\,w_1^2\,\frac{\partial\sigma_0}{\partial w_1} 
                   - \sigma_2\,\frac{\partial\sigma_2}{\partial w_1}\biggr)\,[f(r) + r\,f'(r)] + \ldots\,,
\label{51}
\end{align}
and for $\alpha = u, w, x, y$
\begin{align}
2\,\delta\,\sigma_0\,\frac{\partial f(r)}{\partial \alpha}= &
\,r\,\delta\,\frac{\partial\sigma_0}{\partial \alpha}\,f'(r) 
- w_1^2\,\biggl(\sigma_2\,\frac{\partial\sigma_0}{\partial \alpha}
           - 2\,\sigma_0\,\frac{\partial\sigma_2}{\partial \alpha}\biggr)\,[f''(r) + r\,f'''(r)]\nonumber \\ &
- \sigma_0\,\biggl(2\,w_1^2\,\frac{\partial\sigma_0}{\partial \alpha} 
                   - \sigma_2\,\frac{\partial\sigma_2}{\partial \alpha}\biggr)\,[f(r) + r\,f'(r)] + \ldots\,,
\label{52}
\end{align}
where $\ldots$ denotes inhomogeneous terms presented in Eqs. (\ref{C3},\ref{C5}), and where
\begin{equation}
\delta  = 4\,w_1^2\,\sigma_0 - \sigma_2^2 = -(w_1^2-4\,u^2)\,(w_1^2 - 4\, w^2)\,(w_1^2 - 4\, x^2)\,(w_1^2 - 4\,y^2).
\label{53} 
\end{equation}
The solution for higher order derivatives with respect to $r$ can be represented in the matrix form.
Let us introduce a symbol $\vec f$ to denote
\begin{equation}
\vec f(r) = \left(
\begin{array}{r}
f(r)\\
f'(r)\\
f''(r)\\
f'''(r)
\end{array}\right).
\label{54}
\end{equation}
Then
\begin{equation}
\frac{\partial\vec f}{\partial \alpha} = \hat A_\alpha\,\vec f + \vec F_\alpha,
\label{55}
\end{equation}
where $\hat A_\alpha$ and $F_\alpha$ are presented in Appendix C.
These vector differential equations with respect to 
the nonlinear parameter $\alpha$ can be used 
as recursion relations to express the integral with any powers of
electron distances in terms of the master integral $f(r)$ and it's first three derivatives with respect to $r$. 
These recursions become unstable when $w_1\approx 0$, $\delta\approx 0$ or $\sigma_0\approx 0$.
These cases require a separate analysis. In the next Section we consider recursion relations for $w_1=0$
which has applications for standard molecular integrals. 

The particular case of $\sigma_0=0$ happens for an exchange integral
where $w=u$ and $x=y$. Here, one can use recurrence only for $f'(r)$ to generate
the analytic expressions for derivatives with respect to $w-u$ and $x-y$. They are quite compact
as there will be only three independent parameters, and can be expressed in terms of  $f(r)$ and $f'(r)$.
The derivatives with respect to remaining nonlinear parameters can be calculated afterwards,
by setting $w=u, x=y$.

\section{Recursions for standard molecular integrals}
We describe in this Section the evaluation of standard molecular integrals, where $w_1=0$.
Recursion relations for the  integral $f(r,n)$ in Eq. (\ref{02}) have been presented in Ref. \cite{rec_h2}. 
Here we rederive them using a more compact notation. In particular cases, close to spurious 
singularities,  we  propose to use these recursions
to derive an analytical expression for individual integrals, 
which are then implemented in the numerical codes.

Let us take Eqs. (\ref{12}) and (\ref{14}), and differentiate them
with respect to $w_1$, $n$ and $n-1$ times respectively at $w_1=0$
\begin{align}
&
  r\,\sigma_{00}\,f(r, n) 
+ \sigma_{20}\,f'(r, n) 
+ r\,\sigma_{20}\,f''(r, n) 
+ (n-1)\,n\,r\,\sigma_{02}\,f(r,n-2)
\nonumber \\&
+ (n-1)\,n\,\sigma_{22}\,f'(r,n-2)
+ (n-1)\,n\,r\,\sigma_{22}\,f''(r,n-2) 
+ 2\,(n-1)\,n\,f^{(3)}(r,n-2) 
\nonumber \\&
+ (n-1)\,n\,r\,f^{(4)}(r,n-2) 
+ (n-3)\,(n-2)\,(n-1)\,n\,f'(r,n-4) 
\nonumber \\&
+ (n-3)\,(n-2)\,(n-1)\,n\,r\,f''(r,n-4) 
 = F_{u_1}(r,n)\,,
\label{56}\\[1ex]
&
  \sigma_{00}\,f(r,n)  
+ \sigma_{20}\,f''(r,n) 
+ (n-1)^2\,\sigma_{02}\,f(r,n-2) 
+ (n-1)^2\,\sigma_{22}\,f''(r,n-2)
\nonumber \\& 
+ (n-1)^2\,f^{(4)}(r,n-2) 
+ (n-3)\,(n-2)^2\,(n-1)\,f''(r,n-4) 
= F_{w_1}(r,n-1)\,.
\label{57}
\end{align}
These are two linear equations for three unknowns $f(r, n), f'(r, n), f''(r, n)$.
The third equation is obtained by elimination of $f''(r, n)$ and further differentiation with respect to $r$.
The solution of these three equations for $f(r,n)$ and $f'(r,n)$ are
\begin{eqnarray}
f(r, n)  &=& \frac{1}{\sigma_{00}}\,\bigl[
- (n-2)\,(n-1)\,\sigma_{02}\,f(r,n-2) 
+ (n-1)\,r\,\sigma_{02}\,f'(r, n-2)
\nonumber \\ &&
+ 2\,(n-1)\,\sigma_{22}\,f''(r, n-2)
+ (n-1)\,r\,\sigma_{22}\,f^{(3)}(r, n-2)
\nonumber \\ &&
+ (n-1)\,(n+2)\,f^{(4)}(r, n-2) 
+(n-1)\,r\,f^{(5)}(r, n-2)
\nonumber \\ &&
+ 4\,(n-3)\,(n-2)\,(n-1)\,f''(r, n-4)
+ 2\,(n-3)\,(n-2)\,(n-1)\,r\,f^{(3)}(r, n-4) 
\nonumber \\ &&
+ 2\,F_{w_1}(r,n-1) + r\,F'_{w_1}(r,n-1) - F'_{u_1}(r, n)\bigr] \,, \label{58}\\
f'(r,n) &=& -\frac{1}{\sigma_{20}}\,\bigl[
(n-1)\,r\,\sigma_{02}\,f(r, n-2)  
+ (n-3)\,(n-2)\,(n-1)\,n\,f'(r, n-4)
\nonumber \\ &&
+ (n-1)\,n\,\sigma_{22}\,f'(r, n-2) 
+ 2\,(n-3)\,(n-2)\,(n-1)\,r\,f''(r, n-4)\nonumber \\ &&
+ (n-1)\,r\,\sigma_{22}\,f''(r, n-2)
+ 2\,(n-1)\,n\,f^{(3)}(r, n-2)
+ (n-1)\,r\,f^{(4)}(r, n-2)
\nonumber \\ && 
+ r\,F_{w_1}(r,n-1) - F_{u_1}(r, n)\bigr],
\label{59}
\end{eqnarray}  
where
\begin{equation}
F_{X}(r,n) = (-1)^n\,\frac{\partial^n}{\partial w_1^n}\biggr|_{w_1=0}\, F_{X}(r),
\end{equation}
for $X=w_1,u_1$. Equation (\ref{58}) allows one to obtain integral $f(r,n)$ with an
arbitrary power $n-1$ of $r_{12}$ in terms of $f(r)$, for example
\begin{eqnarray}
f(r,0) &=& f(r),\nonumber\\[1ex]
f(r,1) &=& \frac{r^3}{4}\,h_0(r\,u)\,h_0(r\,w)\,j_0(r\,x)\,j_0(r\,y), \nonumber\\[1ex]
f(r,2) &=& \frac{1}{\sigma_{00}}\bigl[ r\,\sigma_{02}\,f'(r) + 2\,\sigma_{22}\,f''(r) 
           + r\,\sigma_{22}\,f^{(3)}(r) + 4\,f^{(4)}(r) + r\,f^{(5)}(r) \nonumber\\&&
           + 2\,F_{w_1}(r,1) + r\,F'_{w_1}(r,1) - F'_{u_1}(r, 2)\bigr], \nonumber\\[1ex]
f(r,3) &=& \frac{r^5}{24}\,\biggl[ -3\,h_1(r\,u)\,h_1(r\,w)\,j_1(r\,x)\,j_1(r\,y)
          - h_0(r\,u)\,h_0(r\,w)\,j_0(r\,x)\,j_0(r\,y) \nonumber \\ && 
          + h_0(r\,w)\,h_2(r\,u)\,j_0(r\,x)\,j_0(r\,y)
          + h_0(r\,u)\,h_2(r\,w)\,j_0(r\,x)\,j_0(r\,y)  \nonumber \\ && 
          + h_0(r\,u)\,h_0(r\,w)\,j_2(r\,x)\,j_0(r\,y)
          + h_0(r\,u)\,h_0(r\,w)\,j_0(r\,x)\,j_2(r\,y)\biggr],
\label{62}
\end{eqnarray}
where $j_n$ and $h_n$ are modified spherical Bessel functions
\begin{eqnarray}
j_n(x) &=& x^n\,\Bigl(\frac{1}{x}\,\frac{d}{d x}\Bigr)^n\,\frac{\sinh(x)}{x},\nonumber \\
h_n(x) &=& x^n\,\Bigl(\frac{1}{x}\,\frac{d}{d x}\Bigr)^n\,\frac{\exp(-x)}{x}.
\label{72}
\end{eqnarray}
One notices, that $f(r,n)$ for odd $n$ can be expressed in terms of $j_n$ and $h_n$ only,
and their numerical evaluation is straightforward. In contrast, numerical evaluation
of the formulas for even $n$ is much more difficult.
If $\sigma_{20}$ is not small, then derivatives of $f(r)$ can be reduced to $f(r)$ and $f'(r)$ using
\begin{equation}
f^{(n)}(r) = -\frac{1}{r\,\sigma_{20}}\,\Bigl[
\sigma_{00}\,\bigl[(n-2)\,f^{(n-3)}(r) + r\,f^{(n-2)}(r)\bigr]
+ \sigma_{20}\,(n-1)\,f^{(n-1)}(r) - F_{u_1}^{(n-2)}(r)\Bigr]
\label{60}
\end{equation}
If $\sigma_{20}$ is small, then the inverse recursion 
\begin{equation}
f^{(n)}(r) = -\frac{1}{(n+1)\,\sigma_{00}}\,\bigl[
  r\,\sigma_{00}\,f^{(n+1)}(r) 
+ (n+2)\,\sigma_{20}\,f^{(n+2)}(r) 
+ r\,\sigma_{20}\,f^{(n+3)}(r) 
- F_{u_1}^{(n+1)}(r) 
\bigr]
\label{66}
\end{equation}
is stable and can be used to obtain all the derivatives 
including the function $f(r)$ itself. Alternatively $f^{(n)}(r)$ 
can be obtained directly, using the integral form of Eq. (\ref{33}).
If $\sigma_{00}$ is small but not $\sigma_{20}$, then these formulas
can be rewritten using Eq. (\ref{59}) to the form without $\sigma_{00}$ in the denominator, for example
\begin{equation}
f(r,2) =\frac{1}{\sigma_{20}}\,\bigl[\sigma_{02}\,f^{(-2)}(r) - r\,\sigma_{02}\,f^{(-1)}(r) -\sigma_{22}\,f(r)
-r\,\sigma_{22}\,f'(r) - 3\,f''(r) - r\,f'''(r)\bigr]+\ldots\label{63}
\end{equation}
where $f^{(-n)}(r)$ is defined by
\begin{equation}
f^{(-n)}(r) = \int_0^r f^{(-n+1)}(r)\,dr\,, \label{64}
\end{equation}
and can be obtained from  Eqs. (\ref{33}) or (\ref{66}).
It  may lead however to other type of numerical instabilities, 
if the branch point starts at $0$, see Eq. (\ref{21}), and in this case
one can build the explicit table of integrals, as in Ref. \cite{bo_h2,bo_hehp}.
When both $\sigma_{20}\approx 0$ and $\sigma_{00}\approx 0$, then
parameters $x, y$ are small and in this case the Neumann expansion \cite{harris2} can be applied.

What remains are the derivative of $f$ with respect to $\alpha = u,w,x, y$ at $w_1=0$.
We adapt here the derivation of corresponding formulas from Ref. \cite{rec_h2}.
One takes fundamental differential equations in $u_1$ and $\alpha$ at $w_1=0$
\begin{align}
& r\,\sigma_{00}\,f(r) + \sigma_{20}\,f'(r) + r\,\sigma_{20}\,f''(r) = F_{u_1}(r)
\label{67}\\
&\frac{1}{2}\,\frac{\partial\sigma_{00}}{\partial\alpha}\,f(r) 
+ \sigma_{00}\,\frac{\partial f(r)}{\partial\alpha} 
+ \frac{1}{2}\,\frac{\partial\sigma_{20}}{\partial\alpha} f''(r)
+ \sigma_{20}\,\frac{\partial f''(r)}{\partial \alpha} = -F_\alpha(r)
\label{68}
\end{align}
differentiates the first equation with respect to $\alpha$, eliminates
$\partial f''(r)/\partial \alpha$ and differentiate resulting
equation again with respect to $r$. The obtained equation
for the derivative of $f$, using $\partial\sigma_{20}/\partial\alpha = \sigma_{20}/\alpha$ is
\begin{equation}
\frac{\partial f}{\partial \alpha} =
- \frac{1}{2\,\alpha}\,f(r)
+ \frac{r}{2}\biggl(\frac{1}{\sigma_{00}}\, \frac{\partial\sigma_{00}}{\partial\alpha} - \frac{1}{\alpha}\biggr)\,f'(r)
+ \frac{G_\alpha(r)}{\sigma_{00}}\,,
\label{69}
\end{equation}
where 
\begin{equation}
G_\alpha(r) = \Bigl[\Bigl(\frac{1}{2\,\alpha} - \frac{\partial}{\partial\alpha}\Bigr)\,F'_{u_1}(r)
               - \Bigl(2+r\,\frac{d}{d r}\Bigr)\,F_{\alpha}(r) \Bigr]_{w_1=0} \label{70}
\end{equation}
and
\begin{eqnarray}
G_{w}(r) &=& \frac{1}{w}\,\Bigl[
(u\,y - w\,x)\,(u\,x - w\,y)\,(-F_{1+} - F_{2+} + F_{3+} + F_{4+})\nonumber\\ &&
+ (u + w)\,(x - y)\,(u\,w - x\,y)\,(-F_{1+} + F_{2+}) \nonumber\\ &&
+ (u - w)\,(x + y)\,(u\,w - x\,y)\,(-F_{3+} + F_{4+})\Bigr]
\nonumber \\
G_{u}(r) &=& \frac{1}{u}\,\Bigl[
(u\,y - w\,x)\,(u\,x - w\,y)\,(-F_{1+} - F_{2+} + F_{3+} + F_{4+})\nonumber\\ &&
+ (u + w)\,(x - y)\,(u\,w - x\,y)\,(F_{1+} - F_{2+}) \nonumber\\ &&
+ (u - w)\,(x + y)\,(u\,w - x\,y)\,(-F_{3+} + F_{4+})\Bigr]
\nonumber \\
G_{x}(r) &=&
\frac{1}{x}\,\Bigl[
(u\,y - w\,x)\,(u\,x - w\,y)\,(F_{1+} + F_{2+} - F_{3+} - F_{4+})\nonumber\\ &&
+ (u + w)\,(x - y)\,(u\,w - x\,y)\,(-F_{3+} + F_{4+}) \nonumber\\ &&
+ (u - w)\,(x + y)\,(u\,w - x\,y)\,(-F_{1+} + F_{2+})\Bigr]
\nonumber \\ 
G_{y}(r) &=& \frac{1}{y}\,\Bigl[
(u\,y - w\,x)\,(u\,x - w\,y)\,(F_{1+} + F_{2+} - F_{3+} - F_{4+})\nonumber\\ &&
+ (u + w)\,(x - y)\,(u\,w - x\,y)\,(F_{3+} - F_{4+}) \nonumber\\ &&
+ (u - w)\,(x + y)\,(u\,w - x\,y)\,(F_{1+} - F_{2+})\Bigr]\,.
\label{71}
\end{eqnarray}
Here $\sigma_0$ from the denominator can be removed, when necessary, as in the case of $f(r,n)$
by introducing $f^{(-n)}(r)$.
The formula (\ref{69}) allows one to obtain integrals with arbitrary powers of
electronic distances. It requires however,  further investigations of
various numerical instabilities when the denominator approaches $0$.
In particular, $\sigma_{00}$ vanishes for exchange integrals. In this case $w=u, x=y$
and one can generate a table of integrals avoiding recursions,
by Taylor expansion of the fundamental solution in Eq. (\ref{26}) in $w_1$, $w-u$, and $x-y$.
Alternatively, one can take recursion relations presented in Eq. (\ref{59}) and
integrate it once over $r$.

\section{Summary}
The evaluation of two-center two-electron integrals is a difficult task.
Depending on a physical problem different approaches are  
employed. For the calculation of nonrelativistic energies of a two-electron diatomic molecule
the best way is by the use of  Kolos-Wolniewicz basis functions, namely $w_1=0$
and arbitrary nonlinear parameters $u,w,x,y$. The master integral
can be calculated numerically using Eqs. (\ref{33}) and (\ref{35}).
Integrals with powers of interparticle distances can be obtained
by using analytic recursions Eq. (\ref{58}), (\ref{60}), and \ref{69})
followed by numerical evaluation of obtained formulas.

For the calculation of QED effects in a two-electron diatomic molecule, 
the best way is to use the most general basis functions with nonvanishing $w_1,u,w,x,y$.
The master integral can be evaluated according to Eq. (\ref{26}),
and integrals with additional positive powers of interparticle distance 
can be obtained through recursion relations Eq. (\ref{55}). 
Integrals with negative powers can be obtained by numerical integration
with respect to the corresponding nonlinear parameter.
In order to demonstrate correctness of obtained formulas for two-center two-electron integrals,
Table \ref{table2} presents a simple, examplary calculations of the ground electronic
 $\Sigma_g^+$ state of the H$_2$ molecule.
\begin{table}[htb]
\caption{Examplary calculations of the Born-Oppenheimer energy at the equilibrium distance $R=1.4$ 
         of the ground $\Sigma_g^+$ state of H$_2$ molecule, obtained using functions of the form Eq. (\ref{00}),
         with parameters $w_1 = -0.5, u = 1, w = 1, y = 0.125, x = -0.125$ defined in Eqs. (\ref{11}), 
         $N$ is a number of basis functions, and $n_1+n_2+n_3+n_4+n_5 \leq \Omega$.
         The result with $\infty$ is accurate to all digits and was obtained in \cite{kolos, wol, sims,
         nakatsuji, serra, bo_h2}}. 
\label{table2}
\begin{tabular*}{0.3\textwidth}{llw{3.9}}
\hline\hline
\cent{\Omega} & \cent{N} & \cent{E_{\rm BO} (a.u.)} \\
\hline
 0  & 1  & -1.038\,401\,456  \\
 1  & 4  & -1.157\,108\,377  \\
 2  & 13 & -1.172\,368\,483  \\
 3  & 32 & -1.174\,254\,201  \\
$\infty$&& -1.174\,475\,714 \\
\hline\hline
\end{tabular*}
\end{table}

An additional approach to evaluate two-center integrals is by the Taylor expansion in
small internuclear distance $r$. Its applicability has to be verified numerically.
Nevertheless, it serves at present as a simple check of various formulas for recursion relations
and the master integral. The advantage of calculations by Taylor series, is that expansion terms
are quite simple. They are related to the known one-center (helium-like) integrals
and can be obtained recursively (\ref{44}, \ref{49}, \ref{50}). A similar 
Taylor expansion can in principle be obtained for 
many-body three-linked integrals (at most three odd powers of interparticle integrals).
This would open a window for high precision results in many electron diatomic molecules.

\section*{Acknowledgments}
I wish to thank Janek Derezi\'nski for useful information about properties
of Bessel functions. This work was supported by NCN grant 2012/04/A/ST2/00105.

\appendix
\section{Inhomogeneous terms}
The first order differential equation (\ref{07}), satisfied by
the general four body integral, defined by Eq. (3), involves the inhomogeneous term
$P_\alpha$ for $\alpha = u_i, w_i$. They are related to each other by
\begin{eqnarray}
P_{w_1} &=& P(w_1,u_1; w_2, u_2; w_3,u_3) \nonumber \\
       &=& P(w_1,u_1; w_3, u_3; w_2,u_2) \nonumber \\
P_{u_1} &=& P(u_1,w_1; w_2, u_2; u_3,w_3)\nonumber \\
P_{w_2} &=& P(w_2,u_2; w_3, u_3; w_1,u_1)\nonumber \\
P_{u_2} &=& P(u_2,w_2; w_3, u_3; u_1,w_1)\nonumber \\
P_{w_3} &=& P(w_3,u_3; w_1, u_1; w_2,u_2)\nonumber \\
P_{u_3} &=& P(u_3,w_3; u_1, w_1; w_2,u_2) 
\label{A1}
\end{eqnarray}
The expression for $P$ was obtained in \cite{rec_h2} and is the following
\begin{eqnarray}
    & &  P(w_1,u_1;w_2,u_2;w_3,u_3) \nonumber \\
&=& 
\frac{u_1\,w_1\,[(u_1 + w_2)^2 - u_3^2]}{(-u_1 + u_3 - w_2)\,(u_1 + u_3 + w_2)}\,
\ln\biggl[\frac{u_2 + u_3 + w_1}{u_1 + u_2 + w_1 + w_2}\biggr]
\nonumber \\[1ex] &&
+ \frac{u_1\,w_1\,[(u_1 + u_3)^2 - w_2^2]}{(-u_1 - u_3 + w_2)\,(u_1 + u_3 + w_2)}\,
\ln\biggl[\frac{w_1 + w_2 + w_3}{u_1 + u_3 + w_1 + w_3}\biggr] 
\nonumber \\[1ex] &&
 - \frac{u_1^2\,w_1^2 + u_2^2\,w_2^2 - u_3^2\,w_3^2 + w_1\,w_2\,(u_1^2 + u_2^2 - w_3^2)}{(-w_1 - w_2 + w_3)\,(w_1 + w_2 + w_3)}\,
\ln\biggl[\frac{u_1 + u_2 + w_3}{u_1 + u_2 + w_1 + w_2}\biggr] 
\nonumber \\[1ex] &&
- \frac{u_1^2\,w_1^2 - u_2^2\,w_2^2 + u_3^2\,w_3^2 + w_1\,w_3\,(u_1^2 + u_3^2 - w_2^2) }{(-w_1 + w_2 - w_3)\,(w_1 + w_2 + w_3)}\,
\ln\biggl[\frac{u_1 + u_3 + w_2}{u_1 + u_3 + w_1 + w_3}\biggr]
\nonumber \\[1ex] &&
+ \frac{u_2\,(u_2 + w_1)\,(u_1^2 + u_3^2 - w_2^2) - u_3^2\,(u_1^2 + u_2^2 - w_3^2)}{(-u_2 + u_3 - w_1)\,(u_2 + u_3 + w_1)}\,
\ln\biggl[\frac{u_1 + u_3 + w_2}{u_1 + u_2 + w_1 + w_2}\biggr]
\nonumber \\[1ex] &&
+ \frac{u_3\,(u_3 + w_1)\,(u_1^2 + u_2^2 - w_3^2) - u_2^2\,(u_1^2 + u_3^2 - w_2^2) }{(u_2 - u_3 - w_1)\,(u_2 + u_3 + w_1)}\,
\ln\biggl[\frac{u_1 + u_2 + w_3}{u_1 + u_3 + w_1 + w_3}\biggr] 
\nonumber \\[1ex] &&
- \frac{w_1\,[w_2\,(u_1^2 - u_2^2 + w_3^2) + w_3\,(u_1^2 - u_3^2 + w_2^2)]}{(w_1 - w_2 - w_3)\,(w_1 + w_2 + w_3)}\,
\ln\biggl[\frac{u_2 + u_3 + w_1}{u_2 + u_3 + w_2 + w_3}\biggr] 
\nonumber \\[1ex] &&
- \frac{w_1\,[u_2\,(u_1^2 + u_3^2 - w_2^2) + u_3\,(u_1^2 + u_2^2 - w_3^2)]}{(-u_2 - u_3 + w_1)\,(u_2 + u_3 + w_1)}\,
\ln\biggl[\frac{w_1 + w_2 + w_3}{u_2 + u_3 + w_2 + w_3}\biggr]
\label{A2}
\end{eqnarray}
In the case the nonlinear parameter $\alpha$ is the combination of $w_i, u_i$ as in Eq. (\ref{11}),
then $P_\alpha$ is given by
\begin{eqnarray}
P_w &=& P_{w_2}+P_{w_3} \nonumber \\
P_x &=& P_{w_2}-P_{w_3} \nonumber \\
P_u &=& P_{u_2}+P_{u_3} \nonumber \\
P_y &=& P_{u_3}-P_{u_2}
\label{A3}
\end{eqnarray}
The inhomogeneous term in the fundamental differential equation (\ref{12})
is the inverse Laplace transform of $P$
\begin{equation}
F_\alpha = \frac{1}{2\,\pi\,i}\,\int_{-i\,\infty+\epsilon}^{i\,\infty+\epsilon} dt\,e^{t\,r}\,P_\alpha\Bigr|_{u_1=t}
\label{A4}
\end{equation}
In particular, $F_{u_1}$ is given by
\begin{eqnarray}
F_{u_1}(r) &=& \biggl\{
  w_1\,\biggl(\frac{1}{r^2} + \frac{2\,w_1 + u + w - x + y}{r}\biggr)\,e^{-r\,(u + w + w_1 - x + y)} 
\nonumber \\ &&
- w_1\,\biggl(\frac{1}{r^2} + \frac{u + w + x + y}{r}\biggr)\,e^{-r\,(u + w + x + y)} 
\nonumber \\ &&
 + \biggl[\frac{w_1^2}{2}\,(u - w - x + y) + 2\,u\,w\,(x-y) + 2\,x\,y\,(w-u)\biggr]
\nonumber \\ &&
 \times \biggl[{\rm Ei}(-r\,(2\,u + w_1))\,e^{r\,(u - w - x + y)} 
- {\rm Ei}(-r\,(2\,w + w_1))\,e^{-r\,(u - w - x + y)}\biggr]
\nonumber \\ &&
+ \biggl[\frac{w_1^2}{2}\,(u + w + x + y) + 2\,u\,w\,(x+y) + 2\,x\,y\,(u+w)\biggr]
\nonumber \\ &&
\times\biggl[\biggl({\rm Ei}(2\,r\,(x + y)) - {\rm Ei}(-r\,(w_1 - 2\,x)) - {\rm Ei}(-r\,(w_1 - 2\,y))
\nonumber \\ &&
     - \ln\biggl[\frac{(x + y)\,(w_1 + 2\,u)\,(w_1 + 2\,w)}{(u + w)\,(w_1 - 2\,x)\,(w_1 - 2\,y)}\biggr]\biggr)\,e^{-r\,(u + w + x + y)}
   + {\rm Ei}(-2\,r\,(u + w))\,e^{r\,(u + w + x + y)}\biggr]\biggr\}\nonumber \\ &&
+ \biggl\{ x \rightarrow -x, y \rightarrow -y \biggr\} 
\label{A5}
\end{eqnarray}
where ${\rm Ei}$ is the exponential integral function,
and $F_{w_1}$ is
\begin{eqnarray}
F_{w_1}(r) &=&  \biggl\{
 \biggl[\frac{r\,(u + w + x + y)+1}{r^3}\,
\Bigl(2 - \frac{w_1}{2\,u + w_1} - \frac{w_1}{2\,w + w_1} - \frac{w_1}{w_1 - 2\,x} - \frac{w_1}{w_1 - 2\,y}\Bigr)
\nonumber \\ &&
+ \frac{2}{r}\,\biggl(\frac{(u + w)\,(u + x)\,(u + y)}{2\,u + w_1} 
+ \frac{(u + w)\,(w + x)\,(w + y)}{2\,w + w_1} 
\nonumber \\ &&
- \frac{(u + x)\,(w + x)\,(x + y)}{w_1 - 2\,x} 
- \frac{(u + y)\,(w + y)\,(x + y)}{w_1 - 2\,y} - \frac{(u + w + x + y)^2}{2}\biggr)\biggr]
\nonumber \\ &&
\times e^{-r\,(u + w + x + y)} 
\nonumber \\ &&
+ \biggl[\frac{r\,(u + w + w_1 - x + y) + 1}{r^3}\,
\biggl(-2 + \frac{w_1}{2\,u + w_1} + \frac{w_1}{2\,w + w_1} + \frac{w_1}{w_1 - 2\,x} + \frac{w_1}{w_1 + 2\,y}\biggr) 
\nonumber \\ &&
+ \frac{w_1}{r^2} + \frac{1}{r}\,\biggl(w_1^2 + 4\,(x\,y + u\,x + w\,x - u\,y - w\,y - u\,w)  + (w_1 + u + w - x + y)^2
\nonumber \\ &&
- \frac{2\,(u - w)\,(u + x)\,(u - y)}{2\,u + w_1} + \frac{2\,(u - w)\,(w + x)\,(w - y)}{2\,w + w_1}   
\nonumber \\ &&
+ \frac{2\,(u + x)\,(w + x)\,(x + y)}{w_1 - 2\,x} + \frac{2\,(u - y)\,(w - y)\,(-x - y)}{w_1 + 2\,y}\biggr)\biggr]
\,e^{-r\,(u + w + w_1 - x + y)}
\biggr\}
\nonumber \\ &&
+ \biggl\{ x \rightarrow -x, y \rightarrow -y \biggr\} 
\label{A6}
\end{eqnarray}
One notes, that $F_{w_1}$ involves only the exponential and rational functions.

\section{Numerical evaluation of the master integral}

For the numerical evaluation of the master integral in Eq. (\ref{26}),
one should find all singularities on the integration path \cite{fh}. For this
one decomposes $\sigma -\gamma_{i,j}^2$ into products
\begin{eqnarray}
\sigma -\gamma_{0,0}^2 &=& (u_2 - u_3 + w_1)\, (u_2 + u_3 + w_1)\, (t - u_2 - w_3)
\nonumber \\ &&
 (t + u_2 + w_3)\, (w_1 - w_2 + w_3)\, (w_1 + w_2 + w_3) \nonumber \\
\sigma -\gamma_{3,3}^2 &=& (u_2 - u_3 - w_1)\,(u_2 + u_3 + w_1)\,(-t + u_3 + w_2)
\nonumber \\ &&
(t + u_3 + w_2)\,(w_1 + w_2 - w_3)\,(w_1 + w_2 + w_3)\nonumber \\
\sigma -\gamma_{3,1}^2 &=& (t + u_3 - w_2)\, (-t + u_3 + w_2)\, (-t + u_2 - w_3)
\nonumber \\ &&
 (w_1 + w_2 - w_3)\, (t + u_2 + w_3)\, (w_1 - w_2 + w_3)\nonumber \\
\sigma -\gamma_{0,1}^2 &=& (-t + u_3 - w_2)\,(t + u_3 + w_2)\,(t + u_2 - w_3)
\nonumber \\ &&
(w_1 + w_2 - w_3)\,(-t + u_2 + w_3)\,(w_1 - w_2 + w_3)
\label{B1}
\end{eqnarray}
and observes that $\beta_{0,0}$ vanishes on the negative axis at $t=t_2$, $\beta_{3,3}$ at $t=t_1$,
$\beta_{3,1}$ at $t=t_2$, and  $\beta_{0,1}$ at $t=t_1$. All singularities at other points
$t-u_i - w_j$, $t+u_i-w_j$ and $t-u_i+w_j$ cancel out among various $\ln\beta$'s, because
the original integral Eq. (\ref{01}) is finite at these points and $g$ is analytic.
The cancellation can be achieved explicitly, by combining the corresponding $\beta$'s under the common logarithm.
For this one has to consider all possible orderings of $t_i$. 
Because of the symmetries $1 \leftrightarrow 2$ and $A \leftrightarrow B$ one can assume,
without loosing generality, that $t_1>t_2$ and $t_3>t_4$. 
We remain with six possible orderings of $t_i$,
$t_1 \geq t_2 \geq t_3 \geq t_4,\;\;
t_1 \geq t_3 \geq t_2 \geq t_4,\;\;
t_1 \geq t_3 \geq t_4 \geq t_2,\;\;
t_3 \geq t_1 \geq t_2 \geq t_4,\;\;
t_3 \geq t_1 \geq t_4 \geq t_2,\;\; 
t_3 \geq t_4 \geq t_1 \geq t_2$,
and six corresponding representations of the master integral $f$
\begin{eqnarray}
f(r) &=& \biggl(
  \int_{t_1}^{t_2}     \ln|\beta_{0,0}|
+ \int_{t_2}^{t_3}     \ln|\beta_{0,0}\,\beta_{3,3}|
+ \int_{t_3}^{t_4}     \ln|\beta_{0,0}/\beta_{3,1}|
- \int_{t_4}^{-\infty} \ln|\beta_{0,1}\,\beta_{3,1}|
\biggr)\,\frac{e^{t\,r}}{\sqrt{\sigma}}\,dt \nonumber \\ &=& 
\biggl(
  \int_{t_1}^{t_3}     \ln|\beta_{0,0}|
- \int_{t_3}^{t_2}     \ln|\beta_{0,2}|
+ \int_{t_2}^{t_4}     \ln|\beta_{0,0}/\beta_{3,1}|
- \int_{t_4}^{-\infty} \ln|\beta_{0,1}\,\beta_{3,1}|
\biggr)\,\frac{e^{t\,r}}{\sqrt{\sigma}}\,dt \nonumber \\ &=&
\biggl(
  \int_{t_1}^{t_3}     \ln|\beta_{0,0}|
- \int_{t_3}^{t_4}     \ln|\beta_{0,2}|
- \int_{t_4}^{t_2}     \ln|\beta_{0,0}\,\beta_{0,1}\,\beta_{0,2}|
- \int_{t_2}^{-\infty} \ln|\beta_{0,1}\,\beta_{3,1}|
\biggr)\,\frac{e^{t\,r}}{\sqrt{\sigma}}\,dt \nonumber \\ &=&
\biggl(
- \int_{t_3}^{t_1}     \ln|\beta_{0,0}\,\beta_{0,2}|
- \int_{t_1}^{t_2}     \ln|\beta_{0,2}|
+ \int_{t_2}^{t_4}     \ln|\beta_{0,0}/\beta_{3,1}|
- \int_{t_4}^{-\infty} \ln|\beta_{0,1}\,\beta_{3,1}|
\biggr)\,\frac{e^{t\,r}}{\sqrt{\sigma}}\,dt \nonumber \\ &=&
\biggl(
- \int_{t_3}^{t_1}     \ln|\beta_{0,0}\,\beta_{0,2}|
- \int_{t_1}^{t_4}     \ln|\beta_{0,2}|
- \int_{t_4}^{t_2}     \ln|\beta_{0,0}\,\beta_{0,1}\,\beta_{0,2}|
- \int_{t_2}^{-\infty} \ln|\beta_{0,1}\,\beta_{3,1}|
\biggr)\,\frac{e^{t\,r}}{\sqrt{\sigma}}\,dt \nonumber \\ &=&
\biggl(
- \int_{t_3}^{t_4}     \ln|\beta_{0,0}\,\beta_{0,2}|
- \int_{t_4}^{t_1}     \ln|\beta_{0,0}^2\,\beta_{0,1}\,\beta_{0,2}|
- \int_{t_1}^{t_2}     \ln|\beta_{0,0}\,\beta_{0,1}\,\beta_{0,2}|
- \int_{t_2}^{-\infty} \ln|\beta_{0,1}\,\beta_{3,1}|
\biggr)\,\frac{e^{t\,r}}{\sqrt{\sigma}}\,dt \nonumber \\ 
\label{B3}
\end{eqnarray}
In order to eliminate singularities algebraically,
the products and ratios of $\beta$ can be combined together \cite{harris}
\begin{equation}
\frac{\sqrt\sigma-\gamma}{\sqrt\sigma+\gamma}\,\frac{\gamma'-\sqrt\sigma}{\gamma'+\sqrt\sigma} = \frac{\sqrt\sigma-\Gamma}{\sqrt\sigma+\Gamma}
\label{B4}
\end{equation} 
where
\begin{equation}
\Gamma = \frac{\sigma+\gamma\,\gamma'}{\gamma+\gamma'}
\label{B5}
\end{equation}
As a result, no singularity appears on the integration path, which is demonstrated by
the following decompositions
\begin{eqnarray}
\sigma - \Gamma^2(\beta_{0,1}\,\beta_{3,1}) &=&
-(t - u_3 - w_2)\,(t + u_3 - w_2)\,(t - u_3 + w_2)\,(t + u_3 + w_2)\nonumber \\ &&
(t - u_2 - w_3)\,(t + u_2 - w_3)\,(t - u_2 + w_3)\,(t + u_2 + w_3)/(4\,t^2)\nonumber \\
\sigma - \Gamma^2(\beta_{0,0}/\beta_{3,1})  &=&
- (u_2 - u_3 + w_1)\,(u_2 + u_3 + w_1)\,(t - u_3 - w_2)\,(t + u_3 - w_2)\,(t - u_2 - w_3)\nonumber \\ &&
  (w_1 + w_2 - w_3)\,(t - u_2 + w_3)\,(w_1 + w_2 + w_3)/(t - u_2 - w_1 - w_2)^2\nonumber \\
\sigma - \Gamma^2(\beta_{0,0}\,\beta_{0,2}) &=&
-(-u_2 + u_3 + w_1)\,(u_2 + u_3 + w_1)\,(t + u_3 - w_2)\,(t + u_3 + w_2)\,(t - u_2 + w_3)\nonumber \\ &&
(t + u_2 + w_3)\,(w_1 - w_2 + w_3)\,(w_1 + w_2 + w_3)/(t + u_3 + w_1 + w_3)^2 \nonumber \\
\sigma - \Gamma^2(\beta_{0,0}\,\beta_{0,1}) &=&
-(u_2 - u_3 + w_1)\,(u_2 + u_3 + w_1)\,(t - u_3 + w_2)\,(t + u_3 + w_2)\,(t + u_2 - w_3)\nonumber \\ &&
 (w_1 + w_2 - w_3)\,(t + u_2 + w_3)\,(w_1 + w_2 + w_3)/(t + u_2 + w_1 + w_2)^2
\label{B6}
\end{eqnarray}
Table \ref{table1}
presents numerical results for $f$ for several internuclear distances $r$
and nonlinear parameters $w_1, x,y,u$ and $w$.
For $r<0.01$ $f$ is evaluated from the Taylor series up to $r^8$, 
and for $r>0.01$ from the integral representation in Eq. (\ref{B3}).
 
\begin{table}[htb]
\caption{Master integral $f$ for selected values of internuclear distances $r$ and nonlinear parameters}
\label{table1}
\begin{tabular*}{1.0\textwidth}{w{1.2}w{1.2}w{1.2}w{1,2}w{1,2}w{1.19}w{1.19}w{1.19}}
\hline\hline
\cent{w_1} & \cent{u_2} & \cent{w_2} & \cent{u_3} & \cent{w_3} & \cent{f(0.1)\,10^2} & \cent{f(1.0)\,10^3} & \cent{f(10)\,10^{13}} \\
\hline
 2.5  & 2.0  & 1.5  & 1.0 & 0.5 & 1.539\,488\,720\,658\,182    & 3.811\,561\,883\,331\,994     & 2.916\,697\,700\,943\,504 \\
 2.0  & 2.5  & 1.5  & 1.0 & 0.5 & 1.575\,469\,059\,882\,717    & 3.538\,642\,196\,033\,083     & 1.056\,827\,512\,869\,080 \\
 1.5  & 2.0  & 2.5  & 1.0 & 0.5 & 1.592\,898\,118\,308\,067    & 3.359\,218\,711\,378\,032     & 0.983\,486\,526\,526\,378 \\
 1.0  & 2.0  & 1.5  & 2.5 & 0.5 & 1.687\,851\,128\,764\,463    & 3.142\,086\,593\,091\,502     & 0.648\,185\,277\,118\,373 \\
 0.5  & 2.0  & 1.5  & 1.0 & 2.5 & 1.687\,626\,825\,828\,684    & 2.985\,309\,587\,884\,137     & 0.621\,477\,317\,148\,430 \\
-0.5  & 2.0  & 1.5  & 1.0 & 2.5 & 2.285\,510\,252\,707\,772    & 5.843\,903\,698\,492\,676     & 2.026\,400\,131\,640\,827 \\
\hline\hline
\end{tabular*}
\end{table}

\section{Exact form of general recursion relations}
The recursion relations for the general master integral, as given by Eq. (\ref{55}) 
involves the 4 by 4 matrix $A_\alpha$. For $\alpha =  x, y, u, w$, $A$ is the following
\begin{equation}
A_\alpha = \frac{1}{\delta}\left(
\begin{array}{rrrr}
 - \frac{1}{4}\,\frac{\partial\delta}{\partial \alpha} ,& 
 - \frac{r}{4}\,\Bigl(\frac{\partial \delta}{\partial \alpha} - \frac{2\,\delta}{\sigma_0}\,\frac{\partial\sigma_0}{\partial \alpha}\Bigr) ,& 
 - \frac{w_1^2\,\sigma_\alpha}{2\,\sigma_0} ,& 
 - \frac{r\,w_1^2\,\sigma_\alpha}{2\,\sigma_0} \\ 
 \frac{r\,\sigma_\alpha}{2} ,& ,&
 \frac{r}{4}\,\frac{\partial\delta}{\partial \alpha} ,&\\
 \frac{\sigma_\alpha}{2} ,& 
 \frac{r\,\sigma_\alpha}{2} ,& 
 \frac{1}{4}\,\frac{\partial\delta}{\partial \alpha} ,&
 \frac{r}{4}\,\frac{\partial\delta}{\partial \alpha} \\
-\frac{r\,\sigma_0}{4\,w_1^2}\,\frac{\partial\delta}{\partial \alpha} ,&
- \frac{\delta}{2\,w_1^2}\frac{\partial \sigma_2}{\partial \alpha} ,&
- \frac{r\,\delta}{2\,w_1^2}\,\frac{\partial\sigma_2}{\partial \alpha} -\frac{r\,\sigma_\alpha}{2} ,&
\end{array}
\right)
\label{C1}
\end{equation}
where
\begin{equation}
\sigma_\alpha = \frac{\sigma_2}{w_1^2}\,\frac{\partial\,(w_1^2\,\sigma_0)}{\partial \alpha} 
                 - 2\,\sigma_0\,\frac{\partial\sigma_2}{\partial \alpha}
\label{C2}
\end{equation}
and the inhomogeneous term $\vec F_\alpha$ is
\begin{eqnarray}
\vec F_{\alpha} &=& \frac{1}{\delta}\left(
\begin{array}{l}
  - \frac{\sigma_2}{2\,\sigma_0}\,\frac{\partial\sigma_2}{\partial \alpha}\,\frac{\partial}{\partial r} 
  + \Bigl(2\,w_1^2-\frac{\delta}{\sigma_0}\Bigr)\,\frac{\partial^2}{\partial r\partial \alpha}
  + w_1^2\,\frac{\sigma_2}{\sigma_0}\,\frac{\partial^3}{\partial r^3}
\\
  \frac{\partial\sigma_2}{\partial \alpha}
  - \sigma_2\,\frac{\partial}{\partial \alpha}
  - 2\,w_1^2\,\frac{\partial^3}{\partial r^2\,\partial \alpha}
\\
   \frac{\partial\sigma_2}{\partial \alpha}\,\frac{\partial}{\partial r}
 - \sigma_2\,\frac{\partial^2}{\partial r\partial \alpha}
 - 2\,w_1^2\,\frac{\partial^4}{\partial r^3\partial \alpha}
\\
-\frac{\sigma_2}{2\,w_1^2}\,\frac{\partial \sigma_2}{\partial \alpha}
+2\,\sigma_0\,\frac{\partial}{\partial \alpha}
+\sigma_2\,\frac{\partial^3}{\partial r^2\,\partial \alpha}
\end{array}\right)\,F_{u_1}(r) \nonumber \\ &&+
\frac{1}{\delta}\left(
\begin{array}{l}
 2\,\Bigl(w_1^2 - \frac{\delta}{\sigma_0}\Bigr)
+ r\,\Bigl(2\,w_1^2 - \frac{\delta}{\sigma_0}\Bigr)\,\frac{\partial}{\partial r}
+ \frac{5\,w_1^2\,\sigma_2}{\sigma_0}\,\frac{\partial ^2}{\partial r^2} 
+ \frac{r\,w_1^2\,\sigma_2}{\sigma_0}\,\frac{\partial ^3}{\partial r^3} 
\\
- r\,\sigma_2 
- 8\,w_1^2\,\frac{\partial}{\partial r} 
- 2\,r\,w_1^2\,\frac{\partial^2}{\partial r^2} 
\\
- \sigma_2
- r\,\sigma_2\,\frac{\partial}{\partial r} 
- 10\,w_1^2\,\,\frac{\partial^2}{\partial r^2} 
- 2\,r\,w_1^2\,\frac{\partial^3}{\partial r^3} 
\\
  2\,r\,\sigma_0
+ 4\,\sigma_2\,\frac{\partial}{\partial r}
+ r\,\sigma_2\,\frac{\partial^2}{\partial r^2} 
\end{array}\right)\,F_\alpha(r)
\label{C3}
\end{eqnarray}
For $\alpha=w_1$ matrix $A$ is
\begin{equation}
A_{w_1} = \frac{1}{\delta}\left(
\begin{array}{rrrr}
-\frac{1}{4}\,\frac{\partial\delta}{\partial w_1} ,& 
- \frac{r}{4}\Bigl(\frac{\partial \delta}{\partial w_1} - \frac{2\,\delta}{\sigma_0}\,\frac{\partial\sigma_0}{\partial w_1}\Bigr) ,&
-\frac{w_1^2\,\sigma_{w_1}}{2\,\sigma_0},&
-\frac{r\,w_1^2\,\sigma_{w_1}}{2\,\sigma_0} \\
\frac{r\,\sigma_{w_1}}{2},&
- \frac{\delta}{w_1} ,& 
\frac{r}{4}\,\frac{\partial \delta}{\partial w_1} - \frac{r\,\delta}{w_1}, &\\ 
\frac{\sigma_{w_1}}{2},&
\frac{r\,\sigma_{w_1}}{2},&
\frac{1}{4}\,\frac{\partial\delta}{\partial w_1} -\frac{2\,\delta}{w_1}  ,&
\frac{r}{4}\,\frac{\partial \delta}{\partial w_1} - \frac{r\,\delta}{w_1} \\ 
\frac{r\,\sigma_0\,\delta}{w_1^3}-\frac{r\,\sigma_0}{4\,w_1^2}\,\frac{\partial\delta}{\partial w_1},&
\frac{\delta\,\sigma_2}{w_1^3} - \frac{\delta}{2\,w_1^2}\,\frac{\partial\sigma_2}{\partial w_1} ,&
\frac{r\,\delta\,\sigma_2}{w_1^3} - \frac{r\,\delta}{2\,w_1^2}\,\frac{\partial\sigma_2}{\partial w_1}-\frac{r\,\sigma_{w_1}}{2},&
- \frac{\delta}{w_1}
 \end{array}
\right)
\label{C4}
\end{equation}
and the inhomogeneous term
\begin{eqnarray}
\vec F_{w_1} &=& \frac{1}{\delta}\left(
\begin{array}{l}
   \frac{1}{4\,\sigma_0}\,\Bigl(\frac{\partial\delta}{\partial w_1}-4\,w_1^2\,\frac{\partial\sigma_0}{\partial w_1}\Bigr)
+  \Bigl(2\,w_1^2 -\frac{\delta}{\sigma_0}\Bigr)\,\frac{\partial^2}{\partial r\partial w_1}
-  w_1\,\frac{\sigma_2}{\sigma_0}\,\frac{\partial^3}{\partial r^3}
+ w_1^2\,\frac{\sigma_2}{\sigma_0}\,\frac{\partial^4}{\partial r^3\partial w_1}
\\
  \Bigl(\frac{\partial\sigma_2}{\partial w_1} - \frac{\sigma_2}{w_1}\Bigr)
- \sigma_2\,\frac{\partial}{\partial w_1}
+ 2\,w_1\,\frac{\partial^2}{\partial r^2}
- 2\,w_1^2\,\frac{\partial^3}{\partial r^2\partial w_1}
\\
  \Bigl(\frac{\partial\sigma_2}{\partial w_1}-\frac{\sigma_2}{w_1}\Bigr)\,\frac{\partial}{\partial r}
-  \sigma_2\,\frac{\partial ^2}{\partial r\partial w_1}
+ 2\,w_1\,\frac{\partial^3}{\partial r^3}
- 2\,w_1^2\,\frac{\partial^4}{\partial r^3\partial w_1}
\\
- \frac{\delta}{w_1^3}+\frac{1}{4\,w_1^2}\,\frac{\partial\delta}{\partial w_1}-\frac{\partial\sigma_0}{\partial w_1}
+ 2\,\sigma_0\,\frac{\partial}{\partial w_1}
- \frac{\sigma_2}{w_1}\,\frac{\partial^2}{\partial r^2}
+ \sigma_2\,\frac{\partial^3}{\partial r^2\partial w_1}
\end{array}\right) F_{u_1}(r)  \nonumber \\ &&
+\frac{1}{\delta}\,\left(
\begin{array}{l}
 2\,\Bigl(w_1^2-\frac{\delta}{\sigma_0}\Bigr)
+ r\,\Bigl(2\,w_1^2 - \frac{\delta}{\sigma_0}\Bigr)\,\frac{\partial}{\partial r}
+ 5\,w_1^2\,\frac{\sigma_2}{\sigma_0}\,\frac{\partial^2}{\partial r^2}
+ r\,w_1^2\,\frac{\sigma_2}{\sigma_0}\,\frac{\partial^3}{\partial r^3}
\\
- r\,\sigma_2
- 8\,w_1^2\,\frac{\partial}{\partial r}
- 2\,r\,w_1^2\,\frac{\partial^2}{\partial r^2}
\\
- \sigma_2
- r\,\sigma_2\,\frac{\partial}{\partial r}
- 10\,w_1^2\,\frac{\partial^2}{\partial r^2}
- 2\,r\,w_1^2\,\frac{\partial^3}{\partial r^3}
\\
2\,r\,\sigma_0
+ 4\,\sigma_2\,\frac{\partial}{\partial r}
+ r\,\sigma_2\,\frac{\partial^2}{\partial r^2}
\end{array}\right) F_{w_1}(r)
\label{C5}
\end{eqnarray}

\end{document}